\documentclass[conference]{IEEEtran}
\IEEEoverridecommandlockouts

\usepackage{cite}
\usepackage{amsmath,amssymb,amsfonts}
\usepackage{subcaption, svg, tabularx, multirow, pifont}
\usepackage{listings}
\usepackage{algorithmic}
\usepackage{graphicx}
\usepackage{textcomp}
\usepackage{xcolor}
\def\BibTeX{{\rm B\kern-.05em{\sc i\kern-.025em b}\kern-.08em
    T\kern-.1667em\lower.7ex\hbox{E}\kern-.125emX}}
\newcommand{\cmark}{\ding{51}}%
\newcommand{\xmark}{\ding{55}}%

\begin{document}

\title{Enhancing Multilingual ASR for Unseen Languages via Language Embedding Modeling}

\author{\IEEEauthorblockN{1\textsuperscript{st} Shao-Syuan Huang}
\IEEEauthorblockA{\textit{Computer Science and Information Engineering} \\
\textit{National Taiwan University}\\
Taipei, Taiwan \\
b09902131@csie.ntu.edu.tw}
\and
\IEEEauthorblockN{2\textsuperscript{nd} Kuan-Po Huang\ \ 3$^{\text{rd}}$\ Andy T. Liu}
\IEEEauthorblockA{\textit{Computer Science and Information Engineering} \\
\textit{National Taiwan University, AICS ASUS}\\
Taipei, Taiwan \\
\{f09922005, f07942089\}@ntu.edu.tw}
\and
\IEEEauthorblockN{4\textsuperscript{th} Hung-yi Lee}
\IEEEauthorblockA{\textit{Electrical Engineering} \\
\textit{National Taiwan University}\\
Taipei, Taiwan \\
hungyilee@ntu.edu.tw}
}

\maketitle

\begin{abstract}
Multilingual Automatic Speech Recognition (ASR) aims to recognize and transcribe speech from multiple languages within a single system. 
Whisper, one of the most advanced ASR models, excels in this domain by handling 99 languages effectively, leveraging a vast amount of data and incorporating language tags as prefixes to guide the recognition process. 
However, despite its success, Whisper struggles with unseen languages, those not included in its pre-training.
Motivated by the observation that many languages share linguistic characteristics, we propose methods that exploit these relationships to enhance ASR performance on unseen languages. 
Specifically, we introduce a weighted sum method, which computes a weighted sum of the embeddings of language tags, using Whisper’s predicted language probabilities. In addition, we develop a predictor-based approach that refines the weighted sum embedding to more closely approximate the true embedding for unseen languages.
Experimental results demonstrate substantial improvements in ASR performance, both in zero-shot and fine-tuning settings. 
Our proposed methods outperform baseline approaches, providing an effective solution for addressing unseen languages in multilingual ASR.
\end{abstract}


\section{Introduction}
\label{sec:intro}
Automatic Speech Recognition (ASR) is a critical area of machine learning research, with applications such as transcription services. Recent advances in deep learning, combined with the availability of large-scale datasets, have significantly improved ASR systems~\cite{asr_system_0, asr_system_1, asr_system_2, asr_system_3, asr_system_4, asr_system_5}, making them more accurate and robust across a variety of languages and dialects.

If an ASR system aims to recognize and transcribe speech from multiple languages, i.e., multilingual ASR, training a separate ASR model for each language demands significant computational resources and extensive data for every language, making it impractical. 
One of the most advanced ASR models in recent years to address this challenge is Whisper~\cite{whisper}.
Whisper utilizes a large-scale transformer architecture and is trained on vast, diverse datasets, enabling it to excel in multilingual speech recognition tasks. 
Unlike traditional models that focus on a single language or require separate models for different languages, Whisper employs a multitask training approach, incorporating several tasks such as speech-to-text transcription, language identification, multilingual transcription, and timestamp alignment. This method allows the model to learn multiple related tasks simultaneously, improving its ability to generalize across different types of speech data. 
In our work, we focus on Whisper's multilingual ability. 
The multilingual ability of Whisper is achieved by the use of special tokens, such as language tags, as prefixes to prompt Whisper to recognize different languages \cite{10626762, yang2024prompts, peng23d_interspeech, 10389617}.

While Whisper’s extensive training data and multitask format allow it to handle a wide range of languages, it struggles with languages outside the 99 languages it was trained on, referred to as ``unseen languages." 
When applying Whisper to an unseen language, the simplest approach is to manually define a new language tag and corresponding language embedding for fine-tuning. 
This baseline approach has limitations, as introducing a new language tag may not fully utilize the knowledge Whisper has learned for the 99 original languages. 
Recent studies~\cite{related_work_0} have addressed the issue of unseen languages, but they often require training an additional large language model (LLM) to guide Whisper’s decoder for unseen languages, which can be resource-intensive and time-consuming.

Motivated by the fact that many languages share linguistic relationships, we propose methods that exploit these relationships by computing a weighted sum of the existing language tag embeddings.
This approach prompts Whisper to leverage information from other languages rather than training a new language tag.
As demonstrated in Sections~\ref{sec:results}, these methods significantly improve Whisper's ASR performance on unseen languages. 
Building on the success of the weighted sum approach, we introduce a predictor-based method that predicts the ground truth embedding from the weighted-sum embedding. 
This further enhances performance, underscoring the benefits of leveraging linguistic relationships between languages. 
Notably, this method requires training only a simple MLP-based predictor rather than a LLM.
In summary, our contributions are:
\begin{itemize}
    \item We proposed a method that calculates a weighted sum of Whisper's language embeddings based on predicted language probabilities, leveraging linguistic relationships to improve performance on unseen languages.
    \item Based on the weighted sum methods, we further develop a predictor-based approach that refines the weighted sum embedding to approximate the true embedding for unseen languages.
    \item Through some ablation studies, we show that applying the proposed methods in both fine-tuning and inference stages yields superior results, confirming the effectiveness of the approach.
\end{itemize}

The structure of this paper is as follows: in Section~\ref{sec:method}, we introduce our methods for weighted summing language tag embeddings and the predictor-based methods. 
Section~\ref{sec:exps} describes our zero-shot and fine-tuning experiments on unseen languages. 
In Section~\ref{sec:results}, we present the improvements achieved by our methods and offer analysis and ablation studies. 
Finally, Section~\ref{sec:conclusion} concludes this work.

\section{Methodology}
\label{sec:method}
Many languages share common linguistic origins and characteristics. 
For instance, Cantonese, Mandarin, and Taiwanese exhibit similarities in both linguistic structure and acoustic features. 
Building on this observation, we propose leveraging these shared traits to enhance the performance of the Whisper model.
Motivated by this insight, we introduce a method that exploits the relationships between languages to improve model outcomes.

In the following sections, we will present the default and baseline approach of the Whisper model, followed by our proposed weighted sum methods.
\begin{figure}[ht]
    \centering
    \includegraphics[width=7cm]{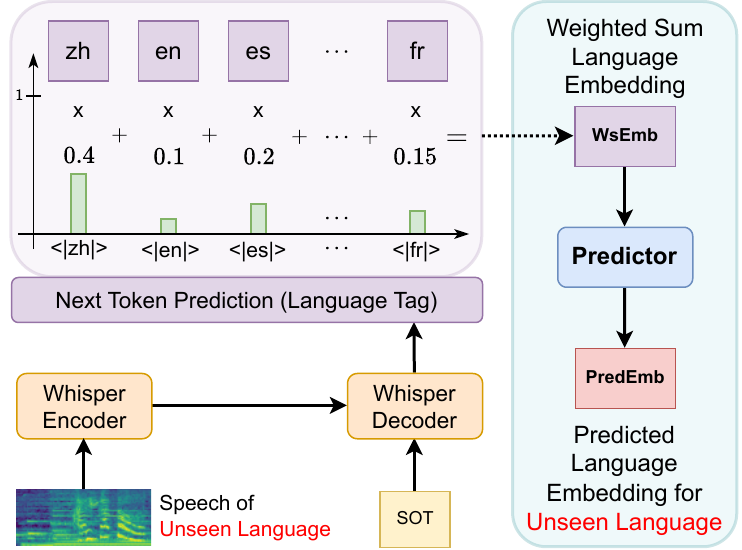}
    \caption{Diagram showing the process of obtaining the weighted sum language embedding, with the yellow boxes indicating the embeddings of individual language tokens.}
    \label{fig:weighted-sum}
\end{figure}

\subsection{Default and Baseline Approach}
\label{sec:baseline_intro}
In Whisper's multitask training framework, a language tag is used to specify the language of a given audio input.
Whisper supports 99 different languages, each represented by a corresponding language tag. 
By assigning a language tag, Whisper is able to recognize various languages without requiring separate models for each one.

For languages not seen during Whisper’s training (denoted as ``Whisper-unseen languages"), if fine-tuning is not feasible, the default approach is to use the tag of the closest language among the 99 languages Whisper was trained on. 
To determine the closest language, we assume the input audio as $x$ and the start-of-text token as $sot$.
Whisper outputs logits for all the possible tokens.
Let $\mathbf{t}=(t_1,t_2,...,t_k,l_1,l_2,...,l_{99})$ represent all tokens, where $l_i$ corresponds to a language tag and $t_i$ represents other tokens.
The logits for the next token $y$ can be formulated as: 
\begin{equation*}
    f(y|x,sot) = \text{Whisper}(x,sot)
\end{equation*}
To obtain the probability distribution for the language tag, the default setting extracts the logits for the language tags and performs a softmax operation:
\begin{equation*}
P(l_i|x,sot) = \frac{exp(f(l_i|x,sot))}{\sum^{99}_{j=1}exp(f(l_j|x,sot))}
\end{equation*}
, where $P(l_i|x,sot)$ represents the probability of the $i$th language tag.
The language tag with the highest probability is selected to represent the unseen language. 
This selected token is then transformed into a token embedding through Whisper’s embedding layer, which we call language embedding. We refer to this approach as the default method, employed in a zero-shot setting where fine-tuning is unavailable.

When fine-tuning is an option, a straightforward way to improve performance is by adding a new language tag to Whisper and fine-tuning the entire model. 
This method is referred to as the baseline method. 
While simple and intuitive, the default and baseline methods overlook potential relationships between different languages due to the pooling operation and the newly added token.

To address this limitation, we propose methods that exploit these inter-language relationships.
\begin{table*}[tbp]
  \center
  \caption{Experimental results on rare languages. The Trainable Embedding column denotes whether the language embedding is trainable during fine-tuning. The Predictor column denotes whether the predictor is used. The Applied Stage shows which stages the weighted sum methods are applied to. The best performance is boldfaced.}
  \label{tab:inference_finetune}
  \resizebox{\linewidth}{!}{
  \begin{tabular}{clcccccc}
    Experiment & \multirow{2}{*}{Method} & Trainable & \multirow{2}{*}{Predictor} & \multicolumn{2}{c}{Applied Stage} & \multicolumn{2}{c}{Metrics} \\
    \cline{5-6}\cline{7-8}
    Setting & & Embedding &  & Fine-tuning & Inference & CER & WER \\
    \hline
    \multirow{3}{*}{Zero-shot} & Default & \xmark & \xmark & \xmark & \xmark & 81.72\% & 139.33\% \\
    & Corpus-wise & \xmark & \xmark & \xmark & \cmark & \textbf{63.66\%} & 132.50\% \\
    & Utterance-wise & \xmark & \xmark & \xmark & \xmark & 76.13\% & \textbf{119.67}\% \\
    \hline
    \multirow{8}{*}{Fine-tuning} & Baseline & \cmark & \xmark & \xmark & \xmark & 28.54\% $\pm$ 2.61\% & 69.97\% $\pm$ 2.71\% \\
    & Corpus-wise & \xmark & \xmark & \cmark & \cmark & 26.09\% $\pm$ 1.53\% & 68.33\% $\pm$ 0.92\% \\
    & Baseline and Corpus-wise & \cmark & \xmark & \xmark & \cmark & 30.31\% $\pm$ 1.76\% & 73.33\% $\pm$ 1.68\% \\
    & Utterance-wise & \xmark & \xmark & \cmark & \cmark & 23.97\% $\pm$ 0.35\% & 65.75\% $\pm$ 1.08\% \\
    & Baseline and Utterance-wise & \cmark & \xmark & \xmark & \cmark & 28.25\% $\pm$ 0.29\% & 72.98\% $\pm$ 2.84\% \\
    & Parameterized Corpus-wise & \cmark & \xmark & \cmark & \cmark & 24.02\% $\pm$ 0.15\% & 62.74\% $\pm$ 0.04\% \\
    & Predictor with Corpus-wise & \xmark & \cmark & \xmark & \xmark & \textbf{22.91\% $\pm$ 0.16\%} & \textbf{61.70\% $\pm$ 0.64\%} \\
    & Predictor with Utterance-wise & \xmark & \cmark & \xmark & \xmark & 23.81\% $\pm$ 0.16\% & 65.66\% $\pm$ 0.34\% \\
  \end{tabular}
  }
\end{table*}
\subsection{Weighted Sum Method}
\label{sec:weighted-sum_intro}
For a Whisper-unseen language, we first obtain the probability distribution $P(l_i|x,sot)$ in the same way as described in the default method (Section~\ref{sec:baseline_intro}). 
However, rather than using the language embedding corresponding to the highest probability language tag, we perform a weighted sum of all 99 language embeddings based on the probability distribution.
Assume the embedding for the $i$th language tag is denoted as $\text{Emb}_i$, the weighted-sum embedding ``$\textit{WsEmb}$" can be computed as:
\begin{equation*}
    \textit{WsEmb} = \sum^{99}_{j=1}P(l_j|x,sot)\text{Emb}_j
\end{equation*}
This process is also illustrated in the left part of Figure~\ref{fig:weighted-sum}.
We refer to this approach as the utterance-wise weighted sum method. In this method, the weighted sum embedding is the language embedding for Whisper’s decoder. 
The goal of this method is to capture the relationships between languages, enabling Whisper to treat the unseen language as a blend of the seen languages, thereby leveraging Whisper’s pre-trained weights.

In the utterance-wise weighted sum method, each utterance in the dataset has its own unique probability distribution for the language tags, leading to a distinct weighted-sum embedding for each utterance.
Alternatively, we propose a variation called the corpus-wise weighted sum method, in which the probability distributions across all utterances are averaged to produce a single, shared language embedding for the entire language corpus.

In multilingual scenarios, each language should have its own corpus-wise weighted-sum embedding when using the corpus-wise weighted sum method. 
The utterance-wise approach is particularly beneficial when it is difficult to determine which instances in the dataset correspond to the same language, making individual embeddings for each utterance a practical solution.
In contrast, the corpus-wise setting mitigates the instability that may 
be caused by the variability in embeddings, as using multiple language embeddings for the same unseen language could introduce inconsistency. 
In such cases, a shared embedding for the entire corpus may offer a more stable and effective representation.

\subsection{Parameterized Weighted Sum Method}
\label{sec:trainable_intro}
In both the utterance-wise and corpus-wise weighted sum methods, the manual modification of the language embedding renders the operation non-differentiable, resulting in fixed weights for the embedding layer. 
A potential limitation of this approach is the static nature of the embedding layer, which may hinder performance improvements.

To address this, we propose a variant called the parameterized corpus-wise weighted sum method. 
In this method, the language embedding is treated as a trainable parameter, initialized using the corpus-wise weighted sum embedding.
This approach effectively transforms the corpus-wise weighted sum method into a trainable version, allowing for further optimization of the language embedding during training.

\subsection{Predictor-based Method}
\label{sec:predictor-based_intro}
If the weighted sum methods described above effectively approximate the true embeddings of an unseen language, it should be possible to simulate this relationship using a straightforward model. 
To explore this, we train a model that uses weighted sum embeddings as input to predict the true language embedding ``\textit{PredEmb}" as shown in Fig. \ref{fig:weighted-sum}.
The predictor is described in the right part of Figure~\ref{fig:weighted-sum}.
Details of the training process are provided in Section~\ref{sec:exps}.

There are two variants of this approach: one utilizes the utterance-wise weighted sum embedding as input, while the other employs the corpus-wise weighted sum embedding. 
We refer to these variants as the predictor with utterance-wise and the predictor with corpus-wise, respectively. 
The predicted true embeddings for the unseen language are then used as the language embedding during fine-tuning and inference.

This method also serves as a verification tool to assess whether the weighted sum embeddings reflect the ground truth embeddings for an unseen language.

\section{Experiment Settings}
\label{sec:exps}
To evaluate our proposed methods, we designed experiments for both zero-shot and fine-tuning settings.
For the data, we utilized the ml-superb~\cite{ml-superb} dataset, which includes a range of languages, including some rare ones. 
We selected Whisper-unseen languages from this dataset and combined them into Whisper-unseen training and test sets.
For the experiments, we employed the Whisper large-v2 model. 
Although Whisper large-v3 is almost identical to large-v2, differing only by the addition of a new language and a change in spectrogram input size, we opted for Whisper large-v2 due to certain known issues\footnote{https://github.com/huggingface/peft/issues/1223}.
To assess performance, we applied standard ASR metrics, including character error rate (CER) and word error rate (WER).

\subsection{Zero-shot Experiments}
\label{sec:zero_shot}
We compare our proposed methods with the default setting of Whisper in zero-shot inference. 
Since fine-tuning is not available in this experiment, neither the parameterized corpus-wise method nor the predictor-based methods are employed. 
In this setting, all Whisper parameters remain fixed. 
The only modification is the language embedding, which is adjusted for the utterance-wise and corpus-wise weighted sum methods.

\subsection{Fine-tuning Experiments}
\label{sec:fine-tune}
Next, we incorporate fine-tuning and compare our proposed methods with the baseline method. 
In this experiment, we include three weighted sum methods: utterance-wise, corpus-wise, and parameterized corpus-wise. 
During fine-tuning, we apply LoRA~\cite{lora} to Whisper, with a rank of 32, alpha set to 64, and dropout with 0.05.
For other training details, we use a learning rate of 4.7e-5 and the AdamW optimizer with a weight decay of 0.02. 
All models are fine-tuned for 5 epochs.

In the case of the utterance-wise and corpus-wise weighted sum methods, the language embedding is manually replaced with the weighted sum embedding during both the fine-tuning and inference stages. 
For the parameterized corpus-wise method, the embedding is initially set to the corpus-wise weighted sum embedding and then fine-tuned along with the model. 
During inference, the trained language embedding is used.

\subsection{Predictor-based Fine-tuning Experiments}
\label{sec:predictor-based_fine-tune}
The predictor-based methods require that the predictor be trained first. 
The objective of the predictor is to estimate the true embedding from the weighted-sum language embedding. 
Since there are no ground truth embeddings for unseen languages, we use Whisper-seen languages to train the model.

We select Whisper-seen languages from the ml-superb dataset as our training and validation sets. 
For training and validation of the predictor, we mask out one seen language at a time and compute the weighted sum embedding (as illustrated in the left part of Figure~\ref{fig:weighted-sum}, but excluding the masked language) from the remaining seen languages. 
In this setup, the weighted-sum embedding serves as the input, while the embedding of the masked language is used as the ground truth.

Using this dataset, we train a 2-layer MLP predictor with the AdamW optimizer (weight decay set to 0.01) and mean squared error as the loss function. 
The learning rate is set to 5e-4, and the hidden size is 1280 (matching the Whisper embedding size). 
This configuration is determined through extensive hyperparameter selection.

With the trained predictor, we can obtain predicted embeddings for each language in the Whisper-unseen training and testing sets. 
These predicted embeddings are then used for fine-tuning and inference of Whisper. 
This method can be seen as a variant of the baseline method, but with language embeddings derived from the predictor.

\section{Results}
\label{sec:results}
Table~\ref{tab:inference_finetune} shows the experimental results on rare languages for the zero-shot setting and the fine-tuned setting. 

\subsection{Zero-shot Performance}
\label{sec:zero-shot_result}
In Table~\ref{tab:inference_finetune}, the comparison between the default method, utterance-wise weighted sum, and corpus-wise weighted sum methods shows that weighted sum methods improve upon the default method. 
Specifically, these methods achieve up to a 22\% reduction in character error rate (CER) and a 14\% reduction in word error rate (WER). 
Although CER and WER remain relatively high due to the zero-shot setting, these improvements indicate that the weighted sum methods could enhance Whisper’s default performance.

\subsection{Fine-tuning Performance}
\label{sec:fine-tune_result}
Referencing the fine-tuning section in Table~\ref{tab:inference_finetune}, the comparison of baseline, utterance-wise, corpus-wise, and parameterized corpus-wise methods shows that all weighted sum methods improve upon the baseline method.
Specifically, these methods achieve up to a 16\% reduction in character error rate (CER) and a 10\% reduction in word error rate (WER).

Thus far, the weighted sum methods prove to be beneficial in both zero-shot and fine-tuning scenarios for Whisper, significantly enhancing Whisper’s ASR performance and surpassing traditional approaches for unseen languages.

\subsection{Predictor-based Performance}
\label{sec:predictor-based_result}
Next, we compare the predictor-based methods with the baseline method and the three weighted sum methods discussed in Section~\ref{sec:fine-tune_result}. 
As shown in Table~\ref{tab:inference_finetune}, the predictor with the corpus-wise setting outperforms the weighted sum methods. 
The predictor-based methods achieve an additional improvement of 20\% in character error rate (CER) and 12\% in word error rate (WER) over the baseline.

The success of the predictor supports our assumption from Section~\ref{sec:predictor-based_intro} that the weighted sum embedding is likely relevant to the true embedding of unseen languages and that this relationship can be effectively simulated by a simple model.

\subsection{Ablation Study}
\label{sec:albation_study}
In Section~\ref{sec:fine-tune}, we applied the weighted sum method during both the fine-tuning and inference stages. 
To further assess the effectiveness of this approach, we provide an ablation study to demonstrate that the method can be beneficial when used in fine-tuning alone.
We fine-tuned the Whisper model using the baseline method and then applied the weighted sum method solely during inference. 
We refer to this approach as ``Baseline and corpus-wise" and ``Baseline and utterance-wise."

In Table~\ref{tab:inference_finetune}, comparing ``Baseline and corpus-wise" with ``Corpus-wise" and ``Baseline and utterance-wise" with ``Utterance-wise" shows that applying the weighted sum method only during inference results in worse performance compared to using it during both fine-tuning and inference. 
These findings suggest that the weighted sum methods provide benefits when applied to both Whisper’s training and inference.

\section{Conclusion}
\label{sec:conclusion}
In this paper, we explored methods to improve Whisper's performance on unseen languages by leveraging linguistic relationships between languages. 
We introduced a weighted sum approach that computes a weighted average of the embeddings of known language tags based on Whisper's predicted probabilities, and a predictor-based method that refines these embeddings. 
Our experiments demonstrated that both methods significantly reduce character error rate (CER) and word error rate (WER) in zero-shot and fine-tuning settings, outperforming baseline approaches. 
These results show that our proposed techniques offer an effective and resource-efficient solution to the challenge of handling unseen languages in multilingual ASR systems.

\bibliographystyle{IEEEbib}
\bibliography{refs}
\vspace{12pt}
\end{document}